\documentclass{article}
\usepackage[dvips]{graphicx}

\begin{document}
\title{Measuring the Leptonic CP Phase in $\nu_{\mu} \to \nu_{\mu}$
Oscillations}
\author{
\sc
{Keiichi Kimura$^1$}
{Akira Takamura$^{1,2}$} \\
\sc
{and} \\
\sc
{Tadashi Yoshikawa$^1$}
\\
\\
{\small \it $^1$Department of Physics, Nagoya University,}
{\small \it Nagoya, 464-8602, Japan}\\
{\small \it $^2$Department of Mathematics,
Toyota National College of Technology}\\
{\small \it Eisei-cho 2-1, Toyota-shi, 471-8525, Japan}}
\date{}
\maketitle

\vspace{-9.5cm}
\begin{flushright}
\end{flushright}
\vspace{7.5cm}

\begin{abstract}
In $\nu_{\mu} \to \nu_{\mu}$ oscillations,
we find that the effect of the CP phase $\delta$
becomes large in the region $E<2$ GeV and $L>2000$ km.
In this region, the change of the probability in this channel
reaches about $0.4$ due to the CP phase effect
beyond our expectation in the case of large 1-3 mixing angle.
Furthermore, the CP phase effect have almost same sign over the
region $E>0.5$ GeV so that one may find the signal of CP violation
by measuring the total rate only.
As an example, we use an experimental setup and demonstrate
that the allowed region is limited
to one by combined analysis of $\nu_e$ and $\nu_{\mu}$ events
although there remain three allowed regions by the analysis
of $\nu_e$ events alone.
\end{abstract}


\maketitle

\section{Introduction}

The measurement of the leptonic CP phase is one of the
most important aims in elementary particle physics.
A lot of investigations for this possibility have been performed,
see \cite{Arafune9703,Cervera,Burguet0103,Minakata,Barger,silver,
Minakata0402,Huber} and the references therein.
In most of these investigations,
$\nu_e \to \nu_{\mu}$ or $\nu_{\mu} \to \nu_e$
oscillations (so-called golden channel \cite{Cervera}) are used
for the measurement of $\delta$ and $\nu_e \to \nu_{\tau}$ oscillations
(silver channel \cite{silver}) are also used partly.
However, there are also $\nu_{\mu} \to \nu_{\mu}$ oscillations
as another channel observed in long baseline experiments.
The measurement of $\delta$ in this channel have been considered
to be difficult because the probability $P_{\mu\mu}$
depends almost only on the $\cos \delta$ term and
the CP dependence disappears in the case of $\delta=90^{\circ}$
and $270^{\circ}$.
However, some recent papers discuss the possibility of solving the parameter
ambiguity
by using this channel \cite{Whisnant0208,Gandhi0411}.
In these analysis, the CP dependence becomes small and the
confirmation of $\delta$ seems to be difficult in actual experiments.
However, if the CP dependence becomes large in this channel,
this analysis is very useful, interesting and cost-effective because
the probability of this channel can be measured at the same time
as $\nu_{\mu} \to \nu_e$ probability without any extra apparatus
and we can obtain another information on the CP phase in
superbeam experiments.
In quark sector, the CP violation was discovered in K physics,
but the determination of the CP phase cannot be performed
by the increase of the data in K physics and needed
the B-factory experiments \cite{Bphysics}.
The measurement of the CP phase in two different channels
is considered to be very important also for the lepton sector.
We obtain the information with a different nature by using different
channels.
Furthermore, this opens the window for a verification of the unitarity
in three generation and exploration of new physics.

In the previous paper, we introduced a new index of the leptonic
CP phase dependence $I_{\rm CP}$
and investigated how the information of $\delta$ can be obtained
through the channel of $\nu_{\mu} \to \nu_e$ oscillation
\cite{Kimura0603}.
In this letter, we use $I_{\rm CP}$ for the channel of
$\nu_{\mu} \to \nu_{\mu}$ oscillation and explore the region in
$E$-$L$ plane where the CP dependence becomes large.
We consider one baseline in this region and investigate the contribution
of the CP phase to $\nu_{\mu} \to \nu_{\mu}$ oscillations in detail.
Finally, we simulate both $\nu_{\mu}$ and $\nu_e$ events by using
an experimental setup and compare the analysis of $\nu_e$ events alone
with the combined analysis of $\nu_{\mu}$ and $\nu_e$ events.

\section{Large CP Dependence in $\nu_{\mu} \to \nu_{\mu}$ Oscillations}

At first, we write the Hamiltonian in matter \cite{MNS} as
\begin{equation}
H = O_{23} \Gamma H'\Gamma^\dagger O_{23}^T
\end{equation}
by factoring out $\theta_{23}$ and $\delta$,
where $O_{23}$ is the rotation matrix in the 2-3 generations
and $\Gamma$ is the phase matrix defined by $\Gamma={\rm diag}
(1,1,e^{i\delta})$.
The reduced Hamiltonian $H^{\prime}$ is given by
\begin{eqnarray}
H'=O_{13}O_{12}{\rm diag}(0,\Delta_{21},\Delta_{31})O_{12}^TO_{13}^T
+ {\rm diag}(a,0,0),
\end{eqnarray}
where $\Delta_{ij}=\Delta m_{ij}^2/(2E)=(m_i^2-m_j^2)/(2E)$,
$a=\sqrt2 G_F N_e\simeq 7.56\times 10^{-5}\cdot \rho Y_e$,
$G_F$ is the Fermi constant, $N_e$ is the electron number density,
$Y_e$ is the electron fraction, $E$ is neutrino energy and $m_i$
is the mass of $\nu_i$.
The oscillation probability for $\nu_{\mu} \to \nu_{\mu}$
is proportional to the $\cos \delta$ and $\cos 2\delta$ in
constant matter profile \cite{Kimura0203}
and can be expressed as
\begin{eqnarray}
P_{\mu\mu}=A\cos \delta+C+D\cos 2\delta \label{4}.
\end{eqnarray}
Here, $A$, $C$ and $D$ are determined by parameters other than $\delta$
and are calculated by
\begin{eqnarray}
A&=&4{\rm Re}[(S_{\mu\mu}^{\prime}c_{23}^2+S_{\tau \tau}^{\prime}s_{23}^2)^*
S_{\mu\tau}^{\prime}]c_{23}s_{23}
\label{eq A}, \\
C&=&|S_{\mu\mu}^{\prime}|^2c_{23}^4+|S_{\tau\tau}^{\prime}|^2s_{23}^4 \\
&&+2(|S_{\mu\tau}^{\prime}|^2+{\rm Re}[S_{\mu\mu}^{\prime
*}S_{\tau\tau}^{\prime}])
c_{23}^2s_{23}^2
\label{eq C}, \\
D&=&2|S_{\mu\tau}^{\prime}|^2c_{23}^2s_{23}^2,
\end{eqnarray}
where $S_{\alpha\beta}^{\prime}=\left[\exp (-iH' L) \right]_{\alpha\beta}$.
In the case of symmetric density profile and $\theta_{23}=45^{\circ}$,
the coefficient $A$ coincides with $A_{e\mu}$, which is the coefficient
of $\cos \delta$ in the $\nu_{\mu} \to \nu_e$ oscillation probability,
except for the sign \cite{Takamura0403}.
Namely, $A$ is simplified as
\begin{eqnarray}
A=-A_{e\mu}=-{\rm Re}[S_{\mu e}^{\prime *}S_{\tau e}^{\prime}].
\label{exactA}
\end{eqnarray}
If we notice the small parameters, $\alpha=\Delta_{21}/\Delta_{31}$
and $\sin \theta_{13}=s_{13}$ and count the order of coefficients,
we obtain $A=O(\alpha s_{13})$, $C=O(1)$ and $D=O(\alpha^2 s_{13}^2)$.
Therefore, this leads to the relation $D\ll A \ll C$ and we can safely
neglect the term including $D$ \cite{Takamura0403}.

Next, we introduce $I_{\rm CP}$ following \cite{Kimura0603}.
Suppose that $P_{max}$ and $P_{min}$ as the maximal and
minimal values when $\delta$ changes from $0^{\circ}$ to $360^{\circ}$.
Then, $I_{\rm CP}$ is defined by
\begin{equation}
I_{\rm CP}=\frac{P_{max}-P_{min}}{P_{max}+P_{min}}
\end{equation}
also in $\nu_{\mu} \to \nu_{\mu}$ oscillation.
As the coefficient $D$ can be neglected,
$P_{\mu\mu}$ is approximated by $P_{\mu\mu}\simeq A\cos \delta+C$.
$P_{max}$ and $P_{min}$ correspond to the two end points.
Namely, the maximal and minimal values are given by
$P_{\mu\mu}\simeq A+C$ or $-A+C$ corresponding to $\delta=0^{\circ}$
or $180^{\circ}$.
So, we obtain
\begin{equation}
I_{\rm CP}\simeq \frac{|A|}{C}.
\end{equation}
Next, let us present the region with large $I_{\rm CP}$ in $E$-$L$ plane.
In order to exclude the region with large $I_{\rm CP}$ due to the
small denominator,
we also show the region with large $|A|$, which is the numerator of
$I_{\rm CP}$.
In fig.1, we assume $\sin^2 2\theta_{13}=0.16$, which is
near the upper bound of the CHOOZ experiment \cite{CHOOZ},
and can be measured in the next generation reactor experiments
\cite{double chooz, kaska}.
If $\theta_{13}$ will be determined by this method without ambiguity,
we can take a large step for the measurement of $\delta$
\cite{Minakata0211}.
We use the best-fit values $\Delta m_{21}^2=7.9\times 10^{-5}{\rm eV}^2$,
$\sin^2 \theta_{12}=0.31$, $\Delta m_{31}^2=2.2\times 10^{-3} {\rm eV}^2$
and
$\sin^2 \theta_{23}=0.50$ \cite{Schwetz0510} for other parameters.
We calculate by using the matter density $\rho=3.3$g/cm$^3$.
In fig.1, black color shows the region with $I_{\rm CP}>30\%$ and
$|A|>20\%$ respectively in left and right figures.
Comparing the two figures, we can see that the region with large $I_{\rm
CP}$
does not overlap that with large $|A|$ above $2$ GeV.
This means that the region of large $I_{\rm CP}$ above $2$ GeV just
indicates
the region where the denominator is small.
On the other hand, both regions overlap over wide region below $2$ GeV.
We found from fig.1
that the CP dependence in $\nu_{\mu} \to \nu_{\mu}$
oscillations becomes large in the region $E<2$ GeV and $L>2000$ km.
We choose a baseline of $L=3000$ km as an example in this region
and investigate the behavior of $A$ and $P_{\mu\mu}$ in detail.
See ref. \cite{future} about other baselines.

\begin{center}
\begin{tabular}{cc}
    \resizebox{64mm}{!}{\includegraphics{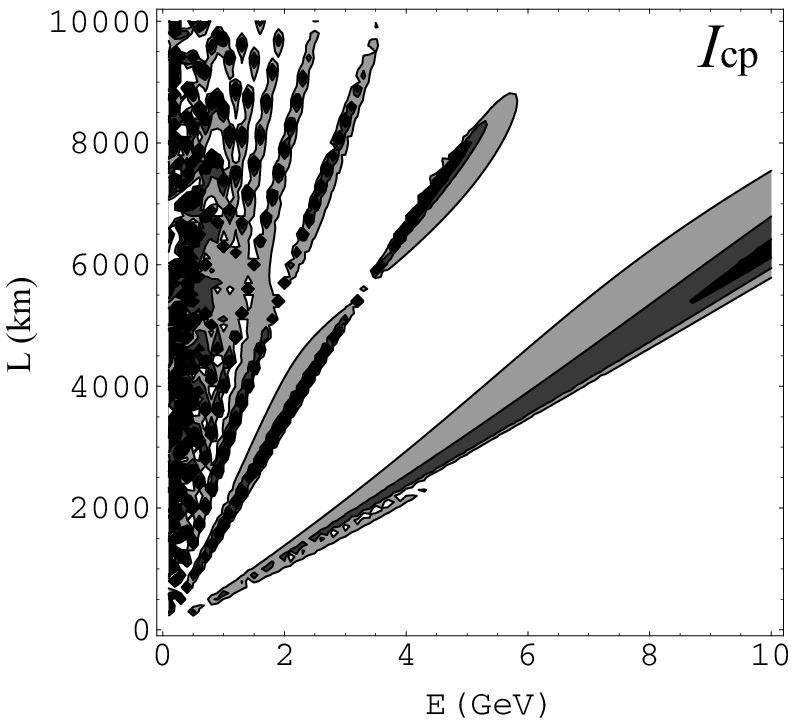}} &
    \resizebox{64mm}{!}{\includegraphics{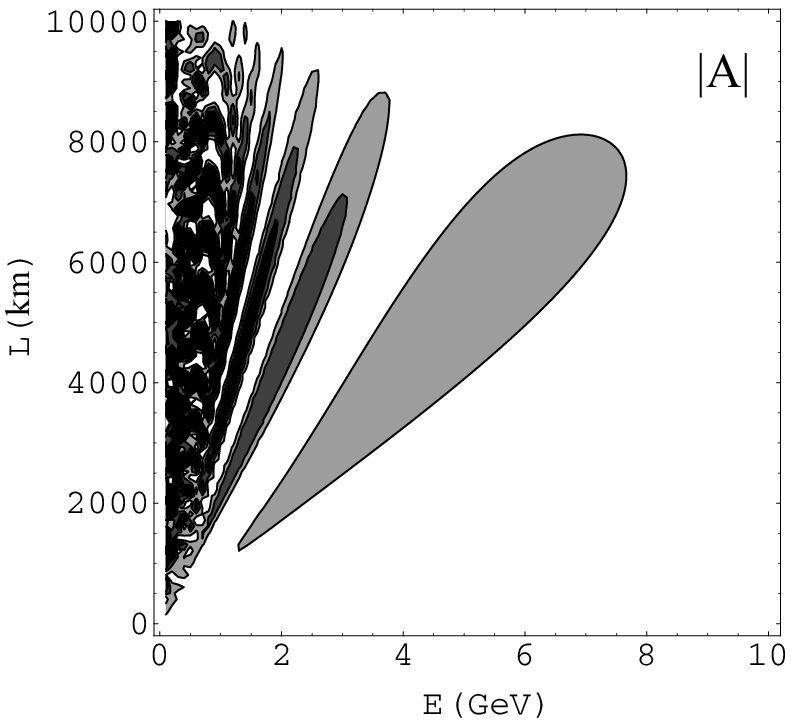}}
\end{tabular}
\vspace{-0.2cm}
\begin{flushleft}
Fig.1. Region with large $I_{\rm CP}$ (Left) and large $|A|$ (Right).
The region with black color has a value larger than $30\%$ for $I_{\rm CP}$
and $20\%$ for $|A|$.
Left and right panels show the region for $I_{\rm CP}$ and $|A|$.
\end{flushleft}
\end{center}

In fig.2, we calculate the energy dependence of $A$ and $P_{\mu\mu}$
at $L=3000$ km by using the exact expressions (\ref{exactA}) and (\ref{4}).
$P_{\mu\mu}$ is plotted in the energy region $E=0.4$-$1.2$ GeV
for the correspondence to the simulation of the number of events.
\begin{center}
\begin{tabular}{ll}
\hspace{-1cm}
\resizebox{75mm}{!}{\includegraphics{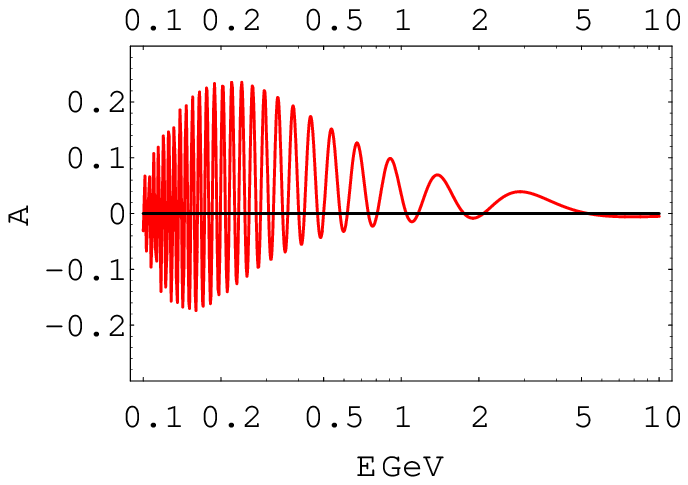}} &
\hspace{-1cm}
\resizebox{75mm}{!}{\includegraphics{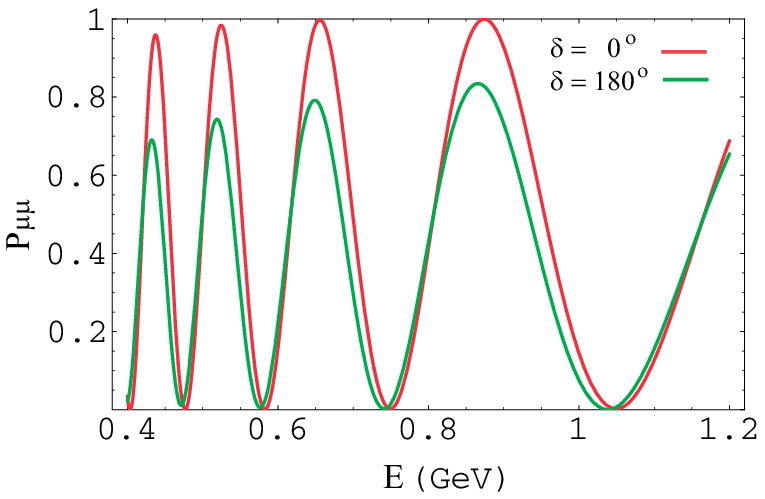}}
\end{tabular}
 \vspace{-0.2cm}
\begin{flushleft}
Fig.2. Energy dependence of $A$ and $P_{\mu\mu}$.
Left figure shows the magnitude of $A$.
In the right figure, the two lines show the oscillation probabilities
with $\delta=0^{\circ}$ and $180^{\circ}$,
respectively.
\end{flushleft}
\end{center}
In fig.2(left), one can see that the maximal value of $A$
reaches almost about $0.2$ and the probabilities for
$\delta=0^{\circ}$ and $180^{\circ}$ have a difference about $0.4$
around $E\simeq 0.2$ GeV.
In addition, the coefficient $A$ is almost positive over the region
$E>0.5$ GeV and we expect a large CP dependence in total
$\nu_{\mu}$ events accumulated in this region.
If the coefficient $A$ oscillates between positive and negative values,
they contribute destructively and the CP phase effect becomes
small in the total events.
Therefore, the fact of $A\geq 0$ in whole region of $E>0.5$ GeV
is very useful for the measurement of the CP phase effect.
This is one of the important points of this paper.

In fig.2 (right), we can see that the probabilities in the case of
$\delta=0^{\circ}$ and $\delta=180^{\circ}$ have a difference
about $0.3$ in the energy region $E=0.4$-$0.6$ GeV.
This can be understood by the magnitude of $A$ for the corresponding
energy region in fig.2 (left).
Up to now, the CP dependence in $\nu_{\mu} \to \nu_{\mu}$ oscillations
is neglected in a lot of works.
Probably, one reason for this is that $P_{\mu\mu}$ depends
only on $\cos \delta$
and the CP dependence disappears in the case of $\delta=90^{\circ}$ and
$270^{\circ}$.
Another reason is
that the analysis satisfying the conditions $E<2$ GeV and $L>2000$ km
is limited.
In this letter, we found that there is a possibility for measuring $\delta$
by using the channel of $\nu_{\mu} \to \nu_{\mu}$ oscillation
under the above conditions.

In order to understand the features in fig.2, let us consider the
approximate formula of $A$.
The expression for (\ref{exactA}) is given by
\begin{equation}
A\simeq -\frac{4\Delta_{21}\Delta_{31}s_{12}c_{12}s_{13}c_{13}}
{\Delta_{\ell}\Delta_{h}}
\sin \frac{\Delta_{\ell}L}{2}
\sin \frac{\Delta_{h}L}{2}
\cos \frac{\Delta_{32}L}{2} \label{A},
\end{equation}
where $\Delta_h=\Delta m^2_h/(2E)$,
$\Delta_{\ell}=\Delta m^2_{\ell}/(2E)$ and
$\Delta m^2_h$ ($\Delta m^2_{\ell}$) is mass squared difference in matter
\cite{Freund,Cervera,Takamura0403}.
The concrete expression for $\Delta_h$ is given by
\begin{equation}
\Delta_{h}=\sqrt{\left(\Delta_{31}\cos 2\theta_{13}-a\right)^2
+\Delta_{31}\sin^2 2\theta_{13}}\simeq \Delta_{31}-a \label{15}.
\end{equation}
The second equality approximately holds in the case of $\sin \theta_{13}\ll
1$.
We also obtain $\Delta_{\ell}$ by the replacements,
$\Delta_{h} \to \Delta_{\ell}$,
$\Delta_{31} \to \Delta_{21}$ and $\theta_{13} \to \theta_{12}$ as
\begin{equation}
\Delta_{\ell}=\sqrt{\left(\Delta_{21}\cos 2\theta_{12}-a\right)^2
+\Delta_{21}\sin^2 2\theta_{12}}\simeq a.
\end{equation}
The second equality approximately holds in the case of
$a\gg \Delta_{21}=\Delta m^2_{21}/2E$.
If we substitute $\rho=3.3$g/cm$^3$ and the electron fraction in the mantle
$Y_e=0.494$, we obtain $a=1.2\times 10^{-4}$ and
$\Delta_{\ell}\simeq a$ becomes a good approximation for
$E\gg \Delta m^2_{21}/2a=0.3$ GeV.
Furthermore, at the baseline $L=3000$ km,
we obtain
\begin{equation}
\sin \frac{\Delta_{\ell}L}{2}\simeq \sin \frac{aL}{2}\simeq 0.8.
\end{equation}
Therefore, the last two terms included in (\ref{A}) are rewritten as
\begin{eqnarray}
2\sin \frac{\Delta_{h}L}{2}\cos \frac{\Delta_{32}L}{2}&\simeq&
2\sin \frac{(\Delta_{31}-a)L}{2}\cos \frac{\Delta_{31}L}{2} \nonumber \\
&&\hspace{-2cm}=\left\{\sin \left(\Delta_{31}L-\frac{aL}{2}\right)-
\sin \frac{aL}{2}\right\} \label{peak},
\end{eqnarray}
and they oscillate within the limits given by
\begin{eqnarray}
-1-\sin \frac{aL}{2}<2\sin \frac{\Delta_{h}L}{2}\cos \frac{\Delta_{32}L}{2}
<1-\sin \frac{aL}{2}.
\end{eqnarray}
Note that this takes value from $-2$ to $0$, namely always negative,
if $\sin \frac{aL}{2}\sim 1$.
As a result, we can roughly estimate $A$ as
\begin{eqnarray}
0\leq A\leq \frac{0.4\Delta_{21}\Delta_{31}}{a(\Delta_{31}-a)}
\simeq \frac{0.4\Delta_{21}}{a}\sim \frac{0.13}{E}
\end{eqnarray}
in $E>0.5$ GeV and can understand the reason for being always
positive in fig.2 (left).
We can also explain that the coefficient $A$ is
inverse proportional to the energy
and decreases according to the increase of the energy.
Furthermore, we find the position of the peak of $A$ as
$2\cdot1.27(\Delta m^2_{31}-aE)L/E=(2n+3/2)\pi$ $(n=0,1,2,\cdots)$
from (\ref{peak}).
If we substitute $\Delta m^2_{31}=2.2\cdot 10^{-3}$eV$^2$ and
$L=3000$km, we obtain the peak energy as
$E=1.27\Delta m^2_{31}L/(2n+7/4)\pi=5.3/(2n+7/4)$.
The peak with highest energy is obtained at $E=3$ GeV
by substituting $n=0$.
Thus, $A$ has large value and is almost positive over the energy
region about $E=0.5$-$5$ GeV.
Here, we also note the $\theta_{13}$ dependence of $A$.
As seen from (\ref{A}), the CP phase effect becomes small
according to the decrease of $\theta_{13}$.

Finally, let us briefly comment on the value of $I_{\rm CP}$ for
$\nu_{\mu} \to \nu_e$ oscillations as preparation of combined
analysis using both $\nu_{\mu}$ and $\nu_e$ events in the next section.
We have investigated how $I_{\rm CP}$ for $\nu_{\mu} \to \nu_e$
oscillations changes according to the values of $E$ and $L$ in the
previous paper \cite{Kimura0603}.
See fig.1 (left) in ref.\cite{Kimura0603}.
As a result, we found that the value of $I_{\rm CP}$ takes nearly maximal 
value
also in the region $E<2$ GeV and $L>2000$ km for $\nu_{\mu} \to \nu_e$
oscillations.
This means that the absolute values of $S_{\tau e}^{\prime}$ and
$S_{\mu e}^{\prime}$ have an same order, namely
the solar term is as large as the atmospheric term in this region.
In the large $I_{CP}$ region, we can expect that
the $\delta$ dependence of the number of events also becomes large.

\section{Estimation of Signal from Leptonic CP Phase}

Next, let us consider an experimental setup
and estimate the CP phase effect.
In order to utilize the $\delta$ dependence of $\nu_{\mu}$ events
in addition to that of $\nu_e$ events,
we would like to choose the experimental setup, which satisfies the
condition $E\simeq 0.5$-$2$ GeV and $L>2000$km obtained in sec.2.
As such an example,
we consider the 4MW beam and the 1Mt water Cherenkov detector,
which are the same used in the JPARC-HyperKamiokande experiment \cite{JHF},
but we take $L=3000$ km as the baseline length.
In fig.3, we numerically calculate the signal of
$\nu_{\mu}$ disappearance (left) and $\nu_e$ appearance (right),
namely the total number of events distinct from the background noise,
obtained in the above experimental setup within the energy
range $E=0.4$-$1.2$ when the CP phase $\delta$ changes from
$0^{\circ}$ to $360^{\circ}$.
We assume here only the neutrino beam data acquisition
for two years and we use the same parameters as in fig.2.
We also give the statistical error within the 2-$\sigma$ level in fig.3.
We use the globes software to perform the numerical calculation
\cite{globes,JHF}.
\begin{center}
\begin{tabular}{cc}
  \resizebox{70mm}{!}{\includegraphics{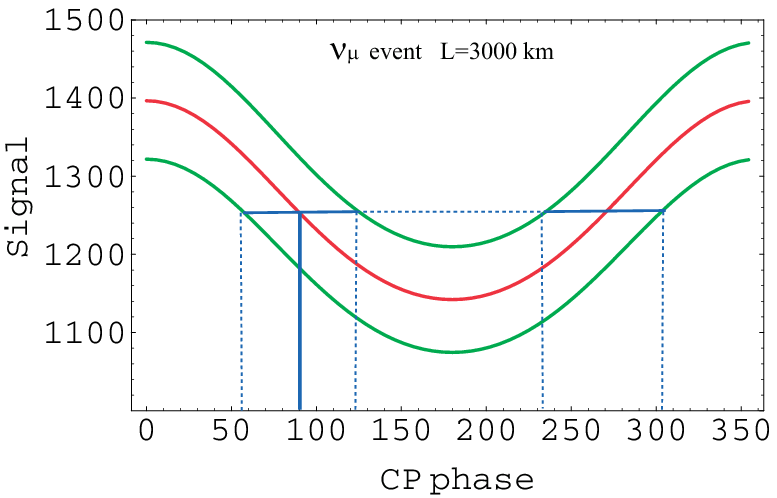}} &
  \resizebox{70mm}{!}{\includegraphics{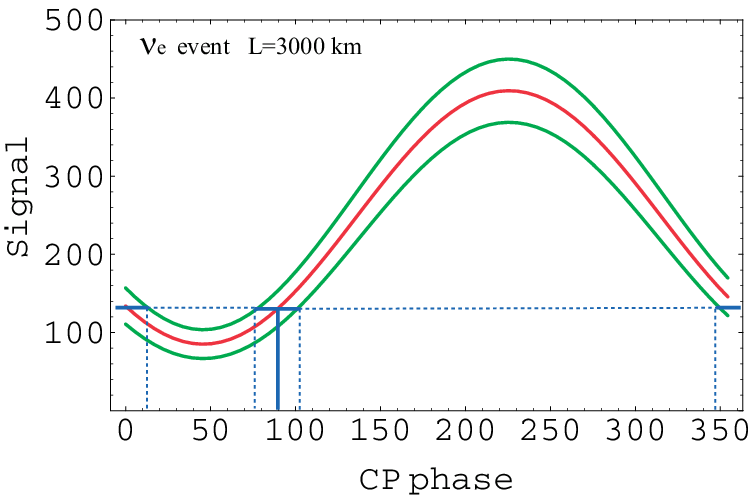}}
\end{tabular}
\vspace{-0.2cm}
\begin{flushleft}
Fig.3. CP dependence of $\nu_{\mu}$ disappearance and $\nu_e$ appearance
signal.
Left and right panels show the $\nu_{\mu}$ and $\nu_e$ signal.
The statistical error is also shown within the 2-$\sigma$ level.
\end{flushleft}
\end{center}
Fig.3 (left) shows that the number of $\nu_{\mu}$ events becomes
maximal around $\delta=0^{\circ}$ and minimal around $\delta=180^{\circ}$
as expected from the oscillation probability in fig.2 (right).
We obtain the number of events between these values in the case of
$\delta=90^{\circ}$ or $270^{\circ}$.
We find that the number of $\nu_{\mu}$ events changes about $250$
due to the CP phase effect.
On the other hand, fig.3 (right) shows that the number of $\nu_e$ events
becomes maximal around $\delta=45^{\circ}$ and minimal around
$\delta=225^{\circ}$.
We also find that the number of $\nu_{e}$ events changes about $300$
due to the CP phase effect.

Next, let us explain the merit of the combined analysis of the total
events in these two channels compared with the analysis of $\nu_e$
events alone.
We assume $\delta=90^{\circ}$ as the true value of the CP phase.
Then, about $120$ $\nu_e$ events are expected and we obtain
the information on the value of $\delta$ as
$0^{\circ}\leq \delta \leq 15^{\circ}$ or
$75^{\circ}\leq \delta \leq 105^{\circ}$ or
$350^{\circ}\leq \delta \leq 360^{\circ}$.
If the standard model is correct and the unitarity holds,
about $1260$ $\nu_{\mu}$ events are expected.
From this result, we obtain the information
$60^{\circ}\leq \delta \leq 130^{\circ}$ or
$235^{\circ}\leq \delta \leq 300^{\circ}$.
If we combine the above information obtained by
the two different channels, the extent of $\delta$
is limited to $75^{\circ}\leq \delta \leq 105^{\circ}$.
Thus, there remain three allowed regions in the analysis
of $\nu_e$ events alone.
On the other hand, one allowed region is chosen in the
combined analysis of two channels.
Furthermore, if the number of $\nu_{\mu}$ events are largely
different from $1260$, this is considered to be the signal
of some new physics beyond the standard model.

In fig.4, we also show the simultaneous fit of $s_{13}$ and
$\delta$ by assuming the true values,
$\sin^2 2\theta_{13}=0.16$ and $\delta=90^{\circ}$.
Other parameters are the same as the values used in fig.3.
We define $\Delta \chi^2_{\nu_e}$ and
$\Delta \chi^2_{\nu_e+\nu_{\mu}}$ as
\begin{eqnarray}
\Delta
\chi^2_{\nu_e}&=&\frac{(N_{\nu_e}-N_{\nu_e}^{true})^2}{N_{\nu_e}^{true}} \\
\Delta \chi^2_{\nu_e+\nu_{\mu}}
&=&\frac{(N_{\nu_e}-N_{\nu_e}^{true})^2}{N_{\nu_e}^{true}}
+\frac{(N_{\nu_{\mu}}-N_{\nu_{\mu}}^{true})^2}{N_{\nu_{\mu}}^{true}},
\end{eqnarray}
where $N_{\nu_i}$ and $N_{\nu_i}^{true}$ represent the number of $\nu_i$
events
calculated by using the test values and true values respectively.
In the left panel, the contours for 1, 2 and 3-$\sigma$ C.L. with
$\Delta \chi^2_{\nu_e}=2.3, 6.18$ and
$11.83$ are drawn as solid, dotted and dashed lines.
In the same way, the contours for $\Delta \chi^2_{\nu_e+\nu_{\mu}}$ are
drawn
in the right panel.
In our analysis, only the statistical error is included for the simplicity.
Here, we also use the globes software to perform
the numerical calculation \cite{globes,JHF}.
\begin{center}
\begin{tabular}{cc}
\hspace{1cm} $\Delta \chi^2_{\nu_e}$ & \hspace{1cm}
$\Delta \chi^2_{\nu_e+\nu_{\mu}}$ \\
  \resizebox{70mm}{!}{\includegraphics{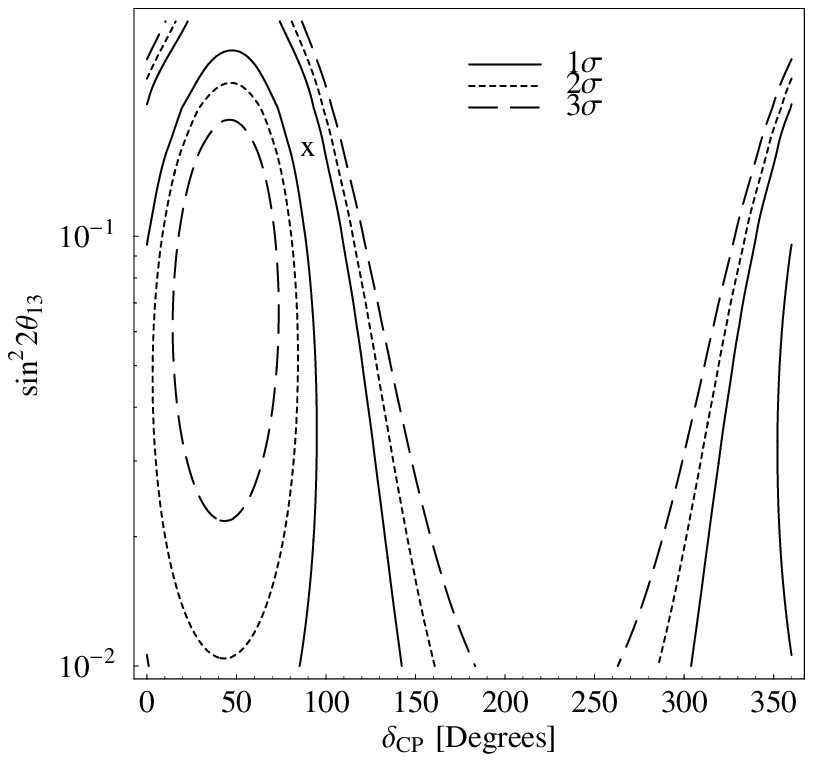}} &
  \resizebox{70mm}{!}{\includegraphics{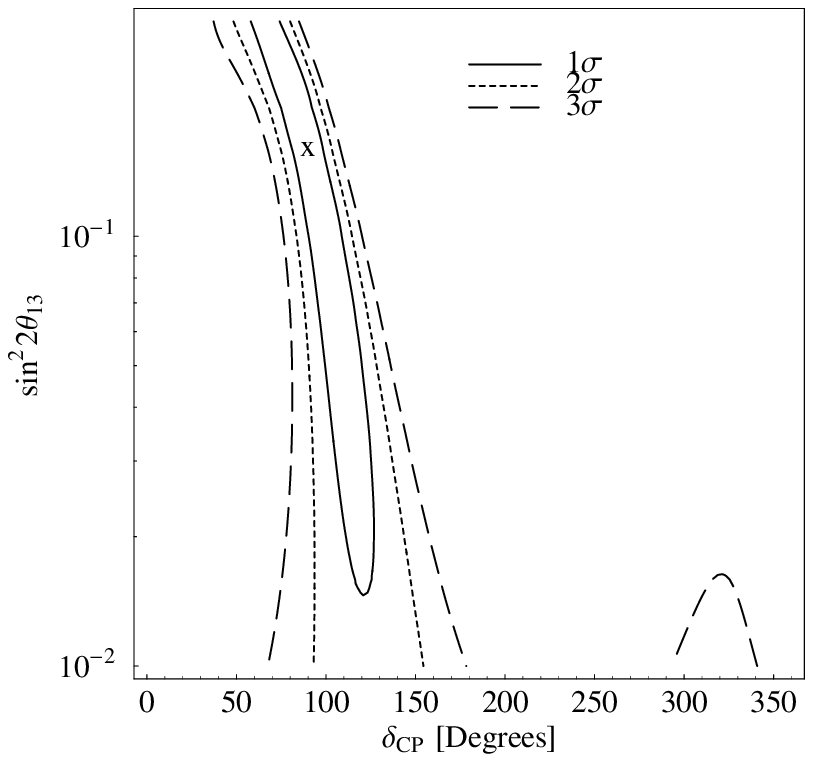}}
\end{tabular}
\vspace{-0.2cm}
\begin{flushleft}
Fig.4. Simultaneous fit of $s_{13}$ and $\delta$ by assuming the
true values, $\sin^2 2\theta_{13}=0.16$ and $\delta=90^{\circ}$.
Left and right panels are drawn assuming
the measurement of $\nu_e$ events alone and
both $\nu_{\mu}$ and $\nu_e$ events.
The true values are represented by x in these figures.
The statistical error is also shown as solid, dotted and dashed
contours for 1,2 and 3-$\sigma$ ($\Delta \chi^2=2.3, 6.18$ and
$11.83$) respectively.
\end{flushleft}
\end{center}
Comparing the left and right panels, we also found that
some allowed regions are excluded by the additional measurement
of $\nu_{\mu}$ events.
Another notable point is that the value of $\Delta \chi^2$ seems to
remain unchanged even in the case for small $\sin^2 2\theta_{13}$.
We can understand this result by the following reason.
The appearance probability for $\nu_{\mu}$ to $\nu_e$ transition
does not change largely because $\theta_{13}$ independent term 
(the solar term)
is comparatively large in such low energy region and
is not so affected by the value of $\theta_{13}$.
Therefore, the value of $\Delta \chi^2$ does not change largely
by the decrease of $\theta_{13}$.

Here, let us emphasize again on the importance of
the simultaneous utilization of the $\nu_{\mu} \to \nu_{\mu}$ and
$\nu_{\mu} \to \nu_e$ oscillations.
This provides the variable possibility
for the measurement of the CP phase in future experiments.
We need to investigate further, including the parameter ambiguity problem
\cite{Burguet0103,Minakata,Barger}, background, systematic error
and spectral dependence \cite{future}.
We also comment on the BNL-HS experiment \cite{BNL}.
In this experiment, $L=2540$ km is taken as the baseline length
and the peak energy of the neutrino beam is planned about $E=1$-$2$ GeV.
This setup satisfy the conditions $E<2$ GeV and $L>2000$km
derived in this letter and is similar to our test experimental setup,
so we expect the observation of the CP phase effect
also in the BNL-HS experiment in the case of large $\theta_{13}$.

\section{Summary and Discussion}

In summary, we have investigated the possibility for the measurement
of the CP phase $\delta$ by using $\nu_{\mu} \to \nu_{\mu}$ oscillations.
The measurement of $\delta$ in this channel have been considered
to be difficult because the probability $P_{\mu\mu}$
depends only on the $\cos \delta$ term and
the CP dependence disappears in the case of $\delta=90^{\circ}$
and $270^{\circ}$.
We have used $I_{\rm CP}$ and $|A|$ (numerator of $I_{\rm CP}$)
for this channel and have found
the CP phase effect becomes large
in the region $L>2000$km and $E<2$ GeV beyond our expectation.
As a result, we have shown that
the difference between the probabilities for
$\delta=0^{\circ}$ and $180^{\circ}$ reaches about $0.4$
in the case of large $\theta_{13}$.
In addition, $A$ is almost positive definite over the region
$E>0.5$ GeV and we expect a large CP dependence in total
$\nu_{\mu}$ events accumulated in this region.
As an example, we have used an experimental setup and
have demonstrated that the allowed region is limited
to one by combined analysis of $\nu_e$ and $\nu_{\mu}$ events
although there remain three allowed regions by the analysis
of $\nu_e$ events alone.
Furthermore, this channel opens the window for a verification of the
unitarity in three generation
and exploration of new physics beyond the standard model.

 \label{sec:summary}

\section*{Acknowledgement}

We would like to thank Prof. Wilfried Wunderlich
(Tokai university) for helpful comments and
advice on English expressions.

\end{document}